\documentclass{aa}  

\usepackage{graphicx}
\usepackage{txfonts,multirow,bigdelim}
\newcommand\kms{km\,s$^{-1}$}

\begin{document}

   \title{Evidence for a chemically differentiated outflow in Mrk~231\thanks{Based on observations with the IRAM Plateau de Bure Interferometer. IRAM is supported by INSU/CNRS (France), MPG (Germany), and IGN (Spain).}}

   \author{J.~E. Lindberg\inst{1,2}\thanks{NASA Postdoctoral Program Fellow.}
          \and
          S. Aalto\inst{1}
          \and
          S. Muller\inst{1}
          \and
          I. Mart\'{i}-Vidal\inst{1}
          \and
          N. Falstad\inst{1}
          \and
          F. Costagliola\inst{1,3}
          \and
          C. Henkel\inst{4,5}
          \and\\
          P. van der Werf\inst{6}
          \and
          S. Garc\'{i}a-Burillo\inst{7}
          \and
          E. Gonz\'{a}lez-Alfonso\inst{8}
          }

   \institute{{Department of Earth and Space Sciences, Chalmers University of Technology, Onsala Observatory,
              439~94 Onsala, Sweden}
              \and
              {NASA Goddard Space Flight Center, Astrochemistry Laboratory, Mail Code 691, 8800 Greenbelt Road, Greenbelt, MD 20771, USA\\}
              \email{johan.lindberg@nasa.gov}
              \and
              {Osservatorio di Radioastronomia (ORA-INAF), Italian ALMA Regional Centre, c/o CNR, Via Gobetti 101, 40129 Bologna, Italy}
              \and
              {Max-Planck-Institut f\"ur Radioastronomie, Auf dem H\"ugel 69, D-53121 Bonn, Germany}
              \and
              {Astronomy Department, King Abdulaziz University, P.O. Box 80203, Jeddah 21589, Saudi Arabia}
              \and
              {Leiden Observatory, Leiden University, PO Box 9513, NL-2300 RA Leiden, The Netherlands}
              \and
              {Observatorio Astron\'{o}mico Nacional (OAN-IGN)-Observatorio de Madrid, Alfonso XII, 3, 28014 Madrid, Spain}
              \and
              {Universidad de Alcal\'{a} de Henares, Departamento de F\'{i}sica, Campus Universitario, 28871 Alcal\'{a} de Henares, Madrid, Spain}
             }

   \date{Received 27 September 2015 / Accepted 14 December 2015}

 
  \abstract
   {}
   {Our goal is to study the chemical composition of the outflows of active galactic nuclei and starburst galaxies.}
   {We obtained high-resolution interferometric observations of HCN and HCO$^+$ $J=1\rightarrow0$ and $J=2\rightarrow1$ of the ultraluminous infrared galaxy Mrk~231 with the IRAM Plateau de Bure Interferometer. We also use previously published observations of HCN and HCO$^+$ $J=1\rightarrow0$ and $J=3\rightarrow2$, and HNC $J=1\rightarrow0$ in the same source.}
   {In the line wings of the HCN, HCO$^+$, and HNC emission, we find that these three molecular species exhibit features at distinct velocities which differ between the species. The features are not consistent with emission lines of other molecular species. Through radiative transfer modelling of the HCN and HCO$^+$ outflow emission we find an average abundance ratio $X(\mathrm{HCN})/X(\mathrm{HCO}^+)\gtrsim1000$. Assuming a clumpy outflow, modelling of the HCN and HCO$^+$ emission produces strongly inconsistent outflow masses.}
   {Both the anti-correlated outflow features of HCN and HCO$^+$ and the different outflow masses calculated from the radiative transfer models of the HCN and HCO$^+$ emission suggest that the outflow is chemically differentiated. The separation between HCN and HCO$^+$ could be an indicator of shock fronts present in the outflow, since the HCN/HCO$^+$ ratio is expected to be elevated in shocked regions. Our result shows that studies of the chemistry in large-scale galactic outflows can be used to better understand the physical properties of these outflows and their effects on the interstellar medium (ISM) in the galaxy.}


   \maketitle
%

\section{Introduction}

Outflows driven by AGNs (active galactic nuclei) and/or starbursts represent a strong and direct mechanism for feedback which may clear central regions of interstellar gas within a few tens of Myr. Many galactic winds and outflows drive out large amounts of molecular gas \citep[see e.g.][]{nakai87,walter02,sakamoto06,tsai09,alatalo11,sturm11,aalto12b,bolatto13} and the ultimate fate of the expelled cold gas is not understood. It is also not clear if the molecular gas is formed in the outflow itself, or if it is carried out from the disk in molecular form. Studying the physical and chemical conditions of the outflowing molecular gas will help us understand the driving mechanism, the origin of the gas and how it is evolving in the wind.

The ultra-luminous infrared galaxy (ULIRG) Mrk~231 ($\log(L_{\mathrm{IR}}/L_\odot)=12.37$), often referred to as the most nearby (175~Mpc) IR quasar (QSO), is a major galaxy merger and hosts both AGN activity and a young, dusty starburst with an extreme star formation rate of $\approx 200 M_{\odot}$~yr$^{-1}$ \citep{taylor99,gallagher02,lipari09}. 

Mrk~231 is well-known for its massive molecular outflow \citep{feruglio10,fischer10}. The outflow rate of molecular gas is estimated to $700M_\odot$~yr$^{-1}$ \citep{feruglio10}, which could empty the reservoir of molecular gas within 10~Myr. \citet{alatalo15} estimates that only $\sim200M_\odot$~yr$^{-1}$ of the molecular gas actually escapes the system, which gives a depletion timescale of $\sim50$~Myr. The high-mass outflow rates originating in an extremely compact nuclear region  \citep[$\sim0.01$~pc;][]{feruglio10,feruglio15} and the wide-angle outflow of neutral atomic gas  \citep[$\sim3$~kpc;][]{rupke11}, both seem to support the notion of an AGN-driven outflow. Furthermore, through far-infrared OH observations \citet{gonzalez14} detected two main outflow components in Mrk~231: one high-velocity component ($\lesssim1500$~\kms) with highly-excited OH emission, indicating that this gas is generated in a compact nuclear region of the galaxy, and a lower-velocity and less excited component ($\lesssim600$~\kms) representing more spatially extended outflowing gas. \citet{teng14} reported that Mrk~231 has an unusually low X-ray luminosity relative to its bolometric luminosity, and argue that this is a result of super-Eddington accretion in the galactic nucleus.

In previous IRAM Plateau de Bure Interferometer (PdBI) observations, \citet{aalto12} found an extremely high HCN/CO 1--0 line ratio (0.3--1) in the Mrk~231 outflow, which suggests that a large fraction of the molecular gas in the outflow is dense ($n\gtrsim10^4$ cm$^{-3}$). Similar HCN/CO line ratios are found in the outflows of M51, suggested to be caused by shocks in the outflow \citep{matsushita15}. HCN has also been detected in the low-velocity molecular
outflow of the Seyfert galaxy NGC~1068 \citep{garcia14}. Recently we reported HCN~3--2/1--0 line ratios in Mrk~231 which suggest that the emission is emerging from gas of $n\approx4$--$5 \times 10^5$ cm$^{-3}$ with a high HCN abundance ($X(\mathrm{HCN})\approx10^{-8}$--$10^{-6}$) and with upper limits of the mass and momentum flux of $4\times10^8M_\odot$ and 12$L_{\rm AGN}/c$, respectively \citep{aalto15b}. The precise structure and driving mechanism of the outflow is, however, still unclear. Among the suggested interpretations are acceleration by the radio jet and boosting of the molecular outflow by an ultrafast outflow (UFO) -- a nuclear semi-relativistic wind \citep{feruglio15}, or (in general) radiation pressure on dust grains \citep[see e.g.][]{murray05}.

In this paper we present IRAM PdBI 3~mm and 2~mm data of the HCN and HCO$^+$ outflow emission in Mrk~231. We also combine these with previously published 3~mm and 1~mm PdBI data \citep{aalto12,aalto15b}. We can show for the first time that the emission from HCN, HCO$^+$, and HNC peaks at different velocities, implying chemical differentiation in the outflow. We interpret this as a result of shocks in the 
outflow and discuss this in relation to its driving mechanism.

\section{Observations}
\label{sec:observations}

\begin{table*}[!htb]
	\centering
	\caption[]{Observational details.}
	\label{tab:obs}
	\begin{tabular}{l l l l l r l}
		\noalign{\smallskip}
		\hline
		\hline
		\noalign{\smallskip}
		Obs. window & Frequency range\tablefootmark{a} & rms\tablefootmark{b} & Obs. date & Synth. beam & P.A.\\
		& [GHz] & [mJy\,(\kms)$^{-1}$] & & [$\arcsec \times \arcsec$] & [\degr] & Array \\
		\noalign{\smallskip}
		\hline
		\noalign{\smallskip}
		3 mm & \phantom{0}85.787--\phantom{0}89.520 & 2.50 & Jan. 2013 & $0.95\times0.80$ & $-8$ & A \\
		3 mm & \phantom{0}87.991--\phantom{0}89.520 & 2.03 & Mar. 2011 \& Jan. 2013 & $1.15\times0.97$ & $33$ & A+B\tablefootmark{c} \\
		2 mm & 175.817--179.554 & 5.21 & May--Oct. 2012 & $2.89 \times 1.99$ & $-62$ & D \\
		\noalign{\smallskip}
		\hline
	\end{tabular}
	\tablefoot{
		\tablefoottext{a}{Rest frequencies assuming the redshift of $z=0.042170$ for Mrk~231 \citep{carilli98}.}
		\tablefoottext{b}{Typical rms per beam per \kms\ in line-free parts of the spectra.}
		\tablefoottext{c}{Combination of the A-array data (above) and B-array data earlier reported by \citet{aalto12}.}
	}
\end{table*}

We have used observations of the HCN 1--0 and 2--1 lines in Mrk~231 carried out with the 6-element IRAM PdBI between summer 2012 and winter 2013. The WideX correlator was configured to cover a bandwidth of 3.6~GHz centered at 84.1~GHz and 170.5~GHz, respectively. We adopt a redshift of $z=0.042170$ (heliocentric frame) for Mrk~231 \citep{carilli98}.
In the final spectra, the 1--0 data were smoothed to $\sim70$~\kms, and the 2--1 data to $\sim40$~\kms.

For the HCN 2--1 tuning, which includes the HCO$^+$ 2--1 line, the configuration was compact (PdBI D array), with a shortest projected baseline of 18~m. This corresponds to a maximum recoverable scale of 12\arcsec. For the HCN 1--0 tuning, which included HCO$^+$ 1--0, the configuration was extended (A array), with a shortest projected baseline of 150~m, corresponding to maximum recoverable scales of 3\arcsec. However, the HCN and HCO$^+$ 1--0 data (including their wide linewings) were combined with data previously obtained in a more compact PdBI configuration \citep[B array;][]{aalto12}, with a maximum recoverable scale of 6\arcsec. This scale is much larger than the extent of the source, in both the line core and the wings, so we do not expect any significant loss of flux. The resulting synthesized beams are between 1--2\arcsec\ depending on the lines. The combination of the two 1--0 datasets was executed using the CASA task \texttt{clean}, where the weights were re-evaluated, normalised by the number of measurement points.

The data reduction was done with the GILDAS/CLIC\footnote{GILDAS is developed by the IRAM institute, Grenoble, France, and can be downloaded from \url{http://www.iram.fr/IRAMFR/GILDAS}} package in a standard way. The bandpass response was calibrated by observing bright radio quasars. The flux calibration was done with MWC349, using the flux model in GILDAS, or the quasar 1150+497. The amplitude and phase gains were derived from the quasar 1150+497. After calibration, the data were exported into CASA\footnote{The Common Astronomy Software Applications package can be acquired from \url{http://casa.nrao.edu/}} for imaging and analysis. We adopt a natural weighting scheme.
Details on the observing dates, resulting rms levels, and synthesized beam sizes are given in Table~\ref{tab:obs}.

In this work, we also use 1~mm PdBI A-array and B-array data of the HCN and HCO$^+$ 3--2 transitions previously reported by \citet{aalto15b}, and 3~mm PdBI B-array observations of the HNC 1--0 transition \citep{aalto12}.

\section{Results}
\label{sec:results}

\subsection{Continuum emission}

The continuum emission detected in the 3~mm A-array data peaks at the position $\alpha=12$:56:14.23, $\delta=+56$:52:25.2 (J2000). The 3~mm (88~GHz) continuum emission in this data set (observed in January 2013) was extracted in channels free from line emission. The continuum level was then estimated through a Gaussian fit of the image plane data. The continuum level was found to be $50.7\pm0.4$~mJy, which is considerably higher than what was found in the older 3~mm (89~GHz) B-array observations; $25.0\pm0.6$~mJy \citep[][observed in March 2011]{aalto12}. During the time period in question, Mrk~231 was undergoing a flare event. Through VLA observations, \citet{reynolds13} found that the 20~GHz continuum flux of Mrk~231 was increased by roughly a factor 2 in this time period. \citet{aalto15b} also found an increase in the 1~mm (256~GHz) continuum flux from 24~mJy to 44~mJy between February 2012 and February 2013, which is consistent with the 20~GHz observations of \citet{reynolds13}.

\subsection{Molecular line emission}

\begin{figure*}[!tb]
    \centering
    \includegraphics{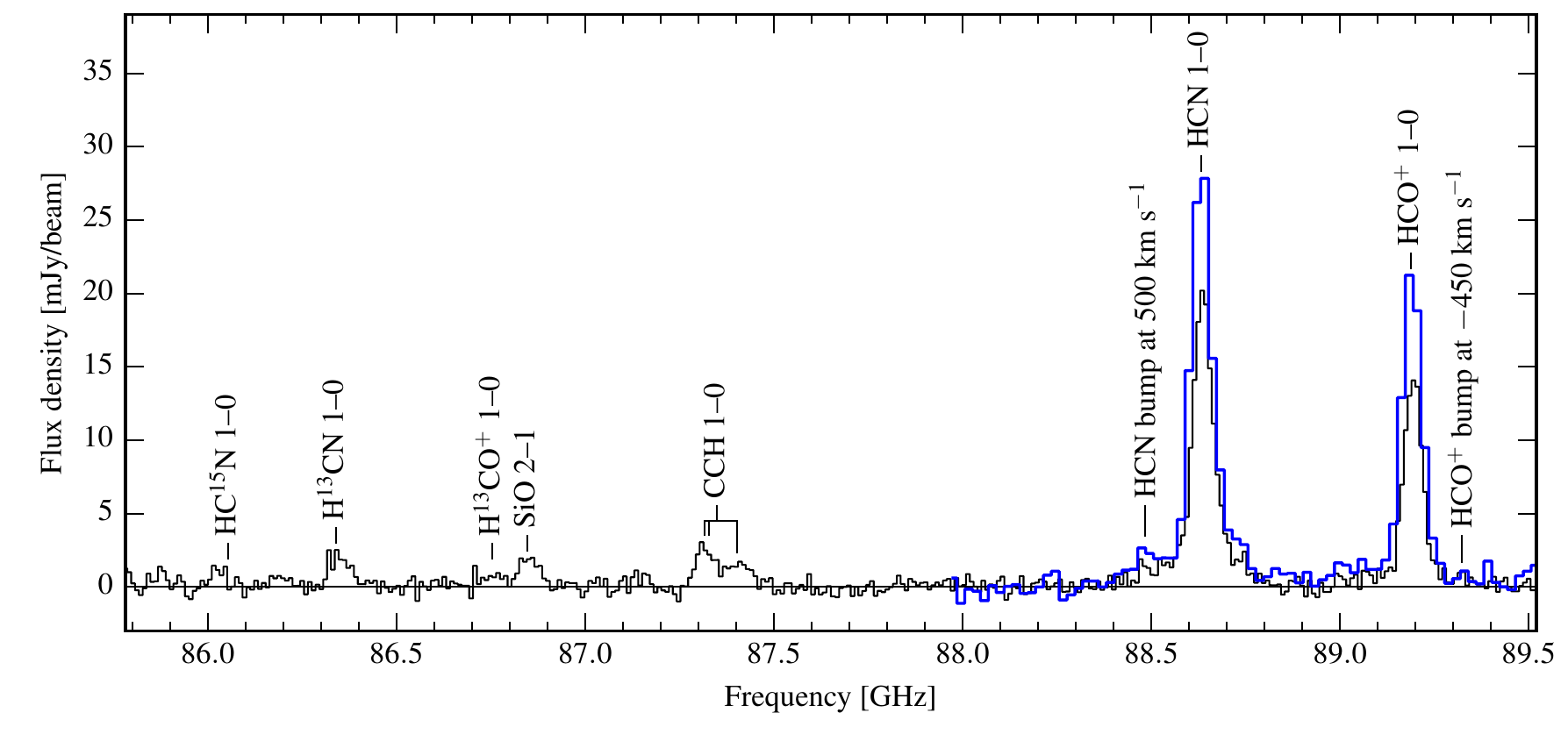}
    \includegraphics{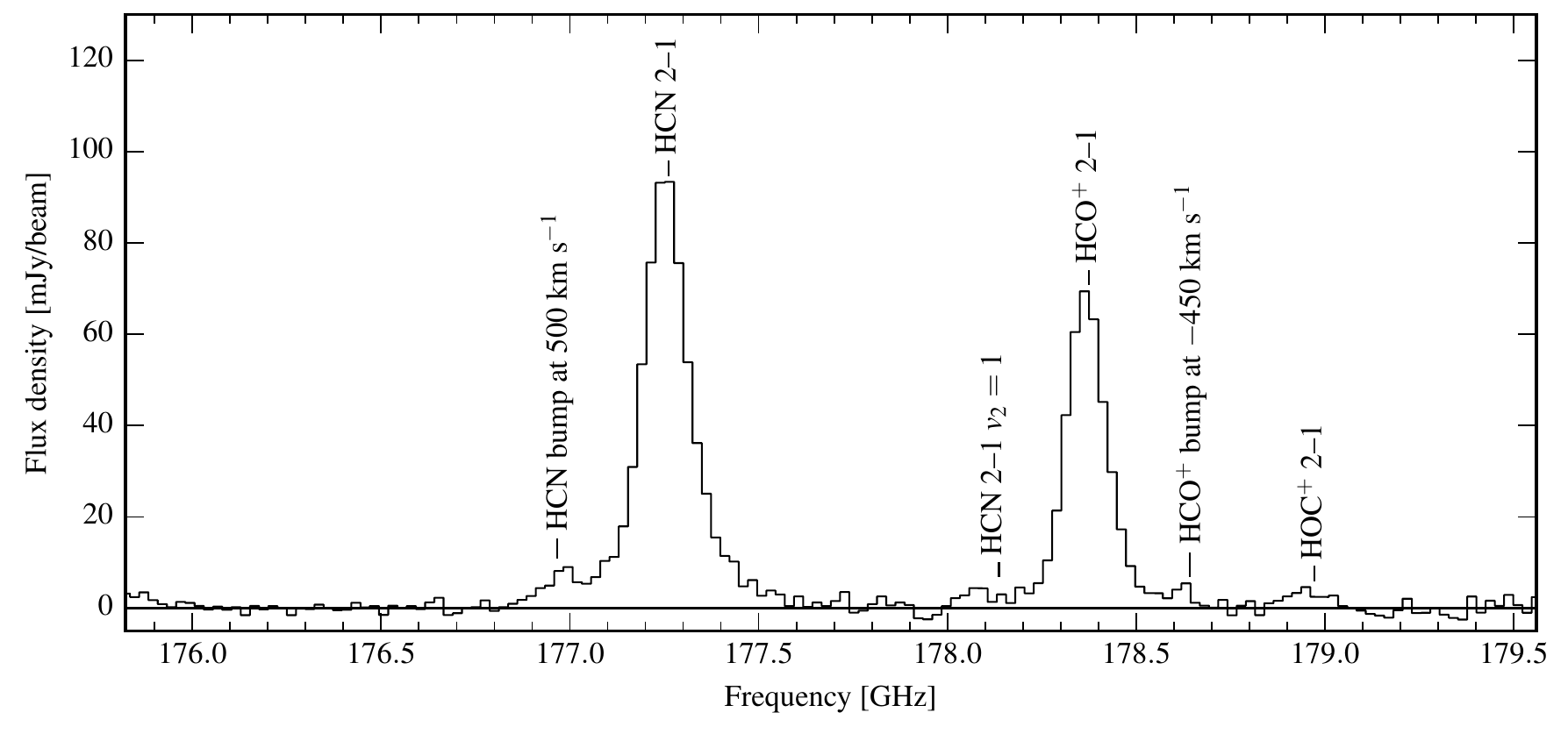}
    \caption{\textit{Top:} 3~mm A-array (black line) and B-array (thick blue line; \citealt{aalto12}) spectra of Mrk~231 towards the central beam. \textit{Bottom:} 2~mm D-array spectrum. The frequency of the HCN 2--1 $v_2=1$ line is marked in the plot, but this line is not significantly detected. \textit{Both figures:} The most prominent outflow bumps of HCN and HCO$^+$ emission are also indicated. The frequency axes of the two spectra are corrected for the redshift of the source ($z=0.042170$). The beam sizes can be found in Table~\ref{tab:obs}.}
    \label{fig:3mm_spectrum}
\end{figure*}

The 3~mm and 2~mm spectra at the peak of the continuum emission are displayed in Fig.~\ref{fig:3mm_spectrum}. In the A-array 3~mm data we detect spectral lines of HCN, HCO$^+$, SiO, CCH, H$^{13}$CN, HC$^{15}$N, and H$^{13}$CO$^+$. In the D-array 2~mm data we detect the HCN, HCO$^+$, and HOC$^+$ 2--1 spectral lines. The HCN and HCO$^+$ 1--0 and 2--1 lines show very wide line wings, with a full width at zero intensity of 1500--2000~\kms. The HCN and HCO$^+$  1--0 lines are also covered in 
PdBI B-array observations \citep{aalto12}, and we combined the two 3~mm data sets to acquire a higher $(u,v)$ coverage and S/N ratio. The HCN and HCO$^+$ lines have somewhat higher flux densities in the B-array data than in the A-array data, owing to some extended emission being resolved out by the extended A-array configuration.

Table~\ref{tab:detections} lists all detected spectral lines in the 3~mm and 2~mm observations, and Table~\ref{tab:linefits} lists extracted fluxes and fitted line parameters. We will focus this paper on the HCN and HCO$^+$ observations.

\begin{table}[!tb]
	\centering
	\caption[]{Detected spectral lines.}
	\label{tab:detections}
	\begin{tabular}{l l l l}
		\noalign{\smallskip}
		\hline
		\hline
		\noalign{\smallskip}
		Species & Transition & Frequency\tablefootmark{a} & $E_{\mathrm{u}}/k$\tablefootmark{a} \\
		& & [GHz] & [K] \\
		\noalign{\smallskip}
		\hline
		\noalign{\smallskip}
		HC$^{15}$N & $J=1\rightarrow0$ & \phantom{0}86		.05497 & \phantom{0}4.13 \\
		H$^{13}$CN & $J=1\rightarrow0$ & \phantom{0}86.33992 & \phantom{0}4.14 \\
		H$^{13}$CO$^+$ & $J=1\rightarrow0$ & \phantom{0}86.75429 & \phantom{0}4.16 \\
		SiO & $J=2\rightarrow1$ & \phantom{0}86.84696 & \phantom{0}6.25 \\
		CCH & $N$ = $1\rightarrow0$\tablefootmark{b} & \phantom{0}87.31690 & \phantom{0}4.19\\
		CCH & $N$ = $1\rightarrow0$\tablefootmark{c} & \phantom{0}87.32859 & \phantom{0}4.19\\
		CCH & $N$ = $1\rightarrow0$\tablefootmark{d} & \phantom{0}87.40199 & \phantom{0}4.20\\
		HCN & $J=1\rightarrow0$ & \phantom{0}88.63160 & \phantom{0}4.25 \\
		HCO$^+$ & $J=1\rightarrow0$ & \phantom{0}89.18852 & \phantom{0}4.28 \\
		\noalign{\smallskip}
		\hline
		\noalign{\smallskip}
		HCN & $J=2\rightarrow1$ & 177.26111 & 12.76 \\
		HCO$^+$ & $J=2\rightarrow1$ & 178.37506 & 12.84 \\
		HOC$^+$ & $J=2\rightarrow1$ & 178.97205 & 12.88 \\
		\noalign{\smallskip}
		\hline
	\end{tabular}
	\tablefoot{
		\tablefoottext{a}{Rest frequencies and energy levels from the CDMS database \citep{cdms}.}
		\tablefootmark{b}{$J$ = $3/2\rightarrow1/2$, $F$ = $2\rightarrow1$.}
		\tablefootmark{c}{$J$ = $3/2\rightarrow1/2$, $F$ = $1\rightarrow0$.}
		\tablefootmark{d}{$J$ = $1/2\rightarrow1/2$, $F$ = $1\rightarrow1$.}
	}
\end{table}

Below, we define the line core emission as any emission between $-250$~\kms\ and 250~\kms\ with respect to the systemic velocity of Mrk~231. Following the \citet{aalto12} definition in Mrk~231, outflow (line wing) emission is any emission between $\pm350$~\kms\ and $\pm990$~\kms.

\subsubsection{HCN}
{\it Outflow:} \, As previously reported \citep{aalto12} and also seen in CO data \citep{cicone12}, the HCN~1--0 line shows emission at velocities $>350$~\kms\ from the line centre which indicates a dense molecular outflow. The red component is more prominent at higher velocities than the blue component. {\it As indicated in Fig.~\ref{fig:3mm_spectrum}, the HCN 1--0 and 2--1 lines show high-velocity bumps, the most prominent at $+500$~\kms.}

\noindent
{\it Line core:} \, We also detect the HCN isotopologues H$^{13}$CN and HC$^{15}$N. H$^{13}$CN was previously detected by \citet{costagliola11}, whereas HC$^{15}$N is detected for the first time in this galaxy. The H$^{13}$CN/HC$^{15}$N line ratio is $2.4\pm0.6$, which is lower than the H$^{13}$CN/HC$^{15}$N line ratio found in NGC~4418 \citep[3.7;][]{sakamoto10}, but higher than in IC~694 \citep[1.2;][]{jiang11} and NGC~4945 \citep[0.96;][]{wang04}.

\citet{aalto15b} detected gas on forbidden velocities westwards of the central peak along the EW-axis by the use of a PV diagram of the HCN~3--2 emission. In this context, forbidden velocities means that this emission does not represent an extension of the rotating disc -- here, the gas is $50$--$100$~\kms\ too fast. This could for instance represent gas in a bar structure, an inflow, or some kind of foreground gas. This emission could also represent low-velocity outflowing gas displaced by shocks. Contrarily, the HCO$^+$ emission lines have no detectable signal at corresponding positions and velocities. \\

\subsubsection{HCO$^+$ and HOC$^+$}
\noindent
{\it Outflow:} \, Like the HCN lines, the HCO$^+$ lines show emission at outflow velocities, although fainter than what is detected in HCN. 
{\it The HCO$^+$ 1--0 and 2--1 lines show high-velocity bumps (Fig.~\ref{fig:3mm_spectrum}) like the HCN lines do, but the most apparent HCO$^+$ bump appears at a lower velocity of approximately $-450$~\kms.}

\noindent
{\it Line core:} \, We also detect the H$^{13}$CO$^+$ 1--0 line and the HOC$^+$ 2--1 line. The HCO$^+$/H$^{13}$CO$^+$ 1--0 ratio is $17\pm4$, whereas the HCN/H$^{13}$CN 1--0 ratio is only $10\pm1$ (both calculated for the A-array data, since the $^{13}$C-species are not covered in the B-array observations), indicating a high optical depth of the HCN line core. We find that the HCO$^+$/HOC$^+$ 2--1 line ratio is only $19\pm3$, which is even lower than what is found in the nearby AGN NGC~1068
\citep[HCO$^+$/HOC$^+$ 1--0 $\approx40$;][]{usero04}, which possibly indicates strong XDR and/or PDR activity in Mrk~231. However, since this is calculated from the possibly optically thick HCO$^+$ 2--1 line, the line ratio might be underestimating the abundance ratio. 

The HCO$^+$ lines are 10--20\% narrower than the HCN lines. No emission at forbidden velocities is found in connection with the line core emission (see also Sec.~\ref{sec:discusscore}).

\begin{table*}[!tb]
	\centering
	\caption[]{Fitted line parameters towards the continuum peak position.}
	\label{tab:linefits}
	\begin{tiny}
		\begin{tabular}{l l r l @{}l l l l l l}
			\noalign{\smallskip}
			\hline
			\hline
			\noalign{\smallskip}
			& & \multicolumn{5}{c}{Core} & \multicolumn{2}{c}{Wings, peak int. intensity}  \\
			Species & Transition & FWHM & Peak flux & & Peak int. intensity\tablefootmark{a} & & Red wing\tablefootmark{b} & Blue wing\tablefootmark{c} & Array \\
			& & [\kms] & [mJy beam$^{-1}$] & & [Jy beam$^{-1}$\,\kms] & & \multicolumn{2}{c}{[Jy beam$^{-1}$\,\kms]} & configuration(s) \\
			\noalign{\smallskip}
			\hline
			\noalign{\smallskip}
			HC$^{15}$N & $J=1\rightarrow0$  & 151 & \phantom{0}1.29 & & $\phantom{0}0.22\pm0.05$ & & ... & ... & A \\
			H$^{13}$CN & $J=1\rightarrow0$  & 226 & \phantom{0}2.17 & & $\phantom{0}0.50\pm0.05$ & & ... & ... & A \\
			H$^{13}$CO$^+$ & $J=1\rightarrow0$  & 258 & \phantom{0}0.77 & & $\phantom{0}0.18\pm0.04$ & & ... & ... & A \\
			SiO & $J=2\rightarrow1$ & 207 & \phantom{0}2.06 & & $\phantom{0}0.45\pm0.05$ & & ... & ... & A \\
			CCH & $N = 1\rightarrow0$\tablefootmark{d} & 172 & \phantom{0}2.73 & \rdelim\}{2}{2.5mm}[] & \multirow{2}{*}{$\phantom{0}0.99\pm0.07$} & & \multirow{2}{*}{...} & \multirow{2}{*}{...} & A \\
			CCH & $N = 1\rightarrow0$\tablefootmark{e} & 314 & \phantom{0}1.58 & & &  & & & A \\
			HCN & $J=1\rightarrow0$ & 217 & 24.5 & & $\phantom{0}5.76\pm0.05$ & & $0.54\pm0.05$ & $0.46\pm0.05$ & A+B \\
			HCO$^+$ & $J=1\rightarrow0$ & 197 & 18.5 & & $\phantom{0}3.95\pm0.05$ & & $0.38\pm0.05$ & $0.21\pm0.05$ & A+B \\
			\noalign{\smallskip}
			\hline
			\noalign{\smallskip}
			HCN &  $J=2\rightarrow1$        & 255 & 90.6 & & $24.09\pm0.14$ & & $1.74\pm0.14$ & $1.18\pm0.14$ & D \\
			HCO$^+$ & $J=2\rightarrow1$     & 215 & 68.0 & & $15.58\pm0.14$ & & $0.78\pm0.14$ & $1.09\pm0.14$\tablefootmark{f} & D \\
			HOC$^+$ & $J=2\rightarrow1$     & 212 & \phantom{0}3.88  & & $\phantom{0}0.82\pm0.12$ & & ... & ... & D \\
			\noalign{\smallskip}
			\hline
		\end{tabular}
	\end{tiny}
	\tablefoot{
		\tablefoottext{a}{Integrated intensity. The HCN and HCO$^+$ lines are integrated between $-250$ and 250~\kms.}
		\tablefoottext{b}{Integrated between 350 and 990~\kms\ towards the continuum peak position.}
		\tablefoottext{c}{Integrated between $-990$ and $-350$~\kms\ towards the continuum peak position.}
		\tablefoottext{d}{Blend of $J$ = $3/2\rightarrow1/2$, $F$ = $2\rightarrow1$ and $J$ = $3/2\rightarrow1/2$, $F$ = $1\rightarrow0$ lines.}
		\tablefoottext{e}{$J$ = $1/2\rightarrow1/2$, $F$ = $1\rightarrow1$ line.}
		\tablefoottext{f}{Blended by the HOC$^+$ $2\rightarrow1$ line.}
	}
\end{table*}

\begin{figure}[!tb]
	\centering
	\includegraphics{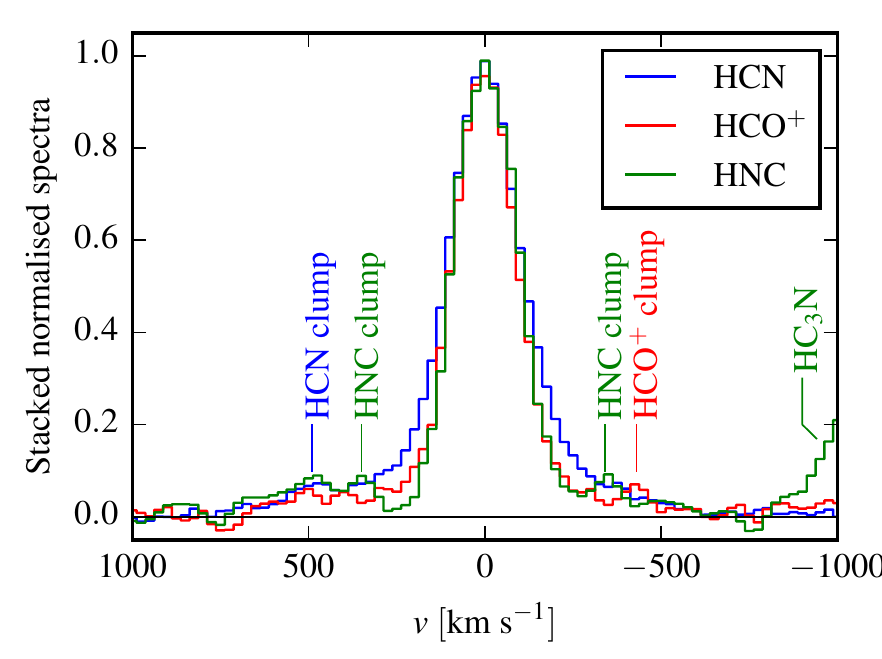}\\
	\caption{The HCN 1--0, 2--1, and 3--2 lines, the HCO$^+$ 1--0 and 2--1 lines, and the HNC 1--0 line, all measured towards the central beam, are here shown normalised, regridded and stacked weighted by the S/N to show the variations in line profile between the species. The HCO$^+$ 3--2 line was not included to avoid contamination from the HCN~3--2 $v_2 = 1$ line. Some of the more important outflow clumps are indicated.}
	\label{fig:compare}
\end{figure}

\section{Discussion}
\label{sec:discussion}

\subsection{Line core emission}
\label{sec:discusscore}

We define line core emission in Mrk~231 as any emission between $-250$~\kms\ and 250~\kms, assuming that this mainly originates in non-outflowing gas.

We normalised the HCN 1--0, 2--1, and 3--2 lines, the HCO$^+$ 1--0 and 2--1 line profiles, and the HNC 1--0 line to unity (the HCO$^+$ 3--2 line was excluded owing to contamination from the HCN~3--2 $v_2 = 1$ line; see below). The normalised HCN and HCO$^+$ lines were then stacked (weighted by their S/N) to allow for a higher-S/N comparison of the molecular emission velocity profiles. The result is shown in Fig.~\ref{fig:compare}, from which it is clear that the HCN emission has a broader core component than the two other species. This extra emission could have the same origin as the forbidden velocity component seen in the HCN~3--2 PV diagram reported by \citet{aalto15b}, representing gas that is not an extension of the rotating disc. However, it could also be explained by high optical depth, possibly even self-absorption, of the HCN line emission (as appears likely from the observed HCN/H$^{13}$CN and HCO$^+$/H$^{13}$CO$^+$ ratios; see above), which causes a slightly broader line profile. Other (U)LIRGs have shown a very high rate of self-absorption of the HCN lines \citep{aalto15a}.

\subsection{Line wing emission}

In the discussion below, we follow the definition of line wings in Mrk~231 from \citet{aalto12}, who define the red line wing to have velocities between 350~\kms\ and 990~\kms, and the blue line wing between $-990$~\kms\ and $-350$~\kms, with respect to the systemic velocity. This emission mainly originates in an almost-face-on outflow. We only detect such wings in the strongest spectral lines: the main isotopologues of HCN, HCO$^+$, and HNC. Outflow line wings have previously also been detected in Mrk~231 through CO observations \citep{cicone12}.

\begin{figure}[!tb]
	\centering
	\includegraphics{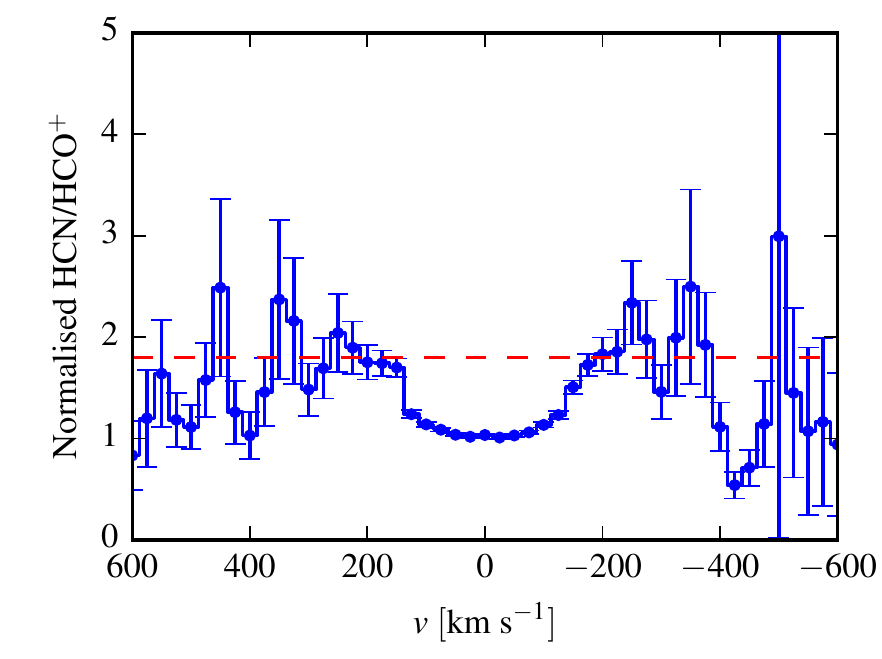}
	\includegraphics{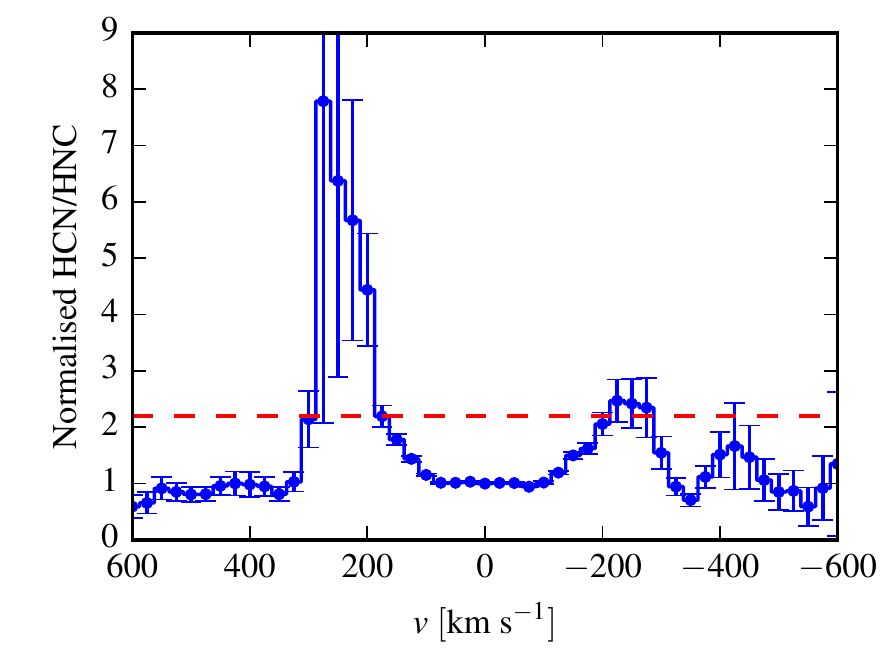}
	\caption{The combined normalised HCN and HCO$^+$ line flux ratios (top panel) and the corresponding HCN and HNC ratios (bottom panel) from Fig.~\ref{fig:compare} are here plotted as a function of velocity. Since the fluxes are normalised, the ratio is per definition 1 at 0~\kms. Error bars are $1\sigma$ (statistical errors only). The red dashed lines show the line core ratio assuming that both species are optically thin towards the edge of the line core (at $\pm200$~\kms). We note that the $y$ axes of the graphs neither show line ratios nor abundance ratios.}
	\label{fig:hcnhcoplus}
\end{figure}

We make a first attempt to study any difference between the HCN, HCO$^+$, and HNC distributions by investigating the ratio of the normalised and stacked HCN, HCO$^+$, and HNC line profiles in Fig.~\ref{fig:compare}. Figure~\ref{fig:hcnhcoplus} shows the HCN/HCO$^+$ and HCN/HNC ratios as a function of velocity. At low ($<150$~\kms) absolute velocities, an increasing optical depth of HCN towards the line centre becomes apparent, but at higher velocities we notice a highly variable and symmetric HCN/HCO$^+$ ratio, oscillating between 1 and 2.5 times the ratio at 0~\kms. The HCN/HNC ratio also shows considerable variation between 1 and $\sim8$ times the ratio at 0~\kms.

\subsubsection{Outflow clumps}

\begin{figure}[!tb]
	\centering
	\includegraphics{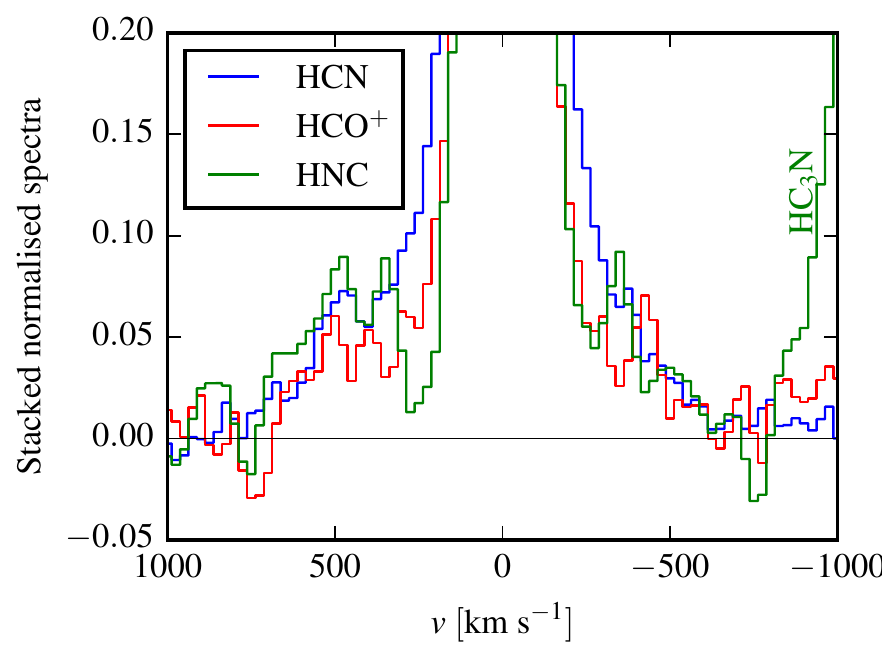}
	\caption{A zoom-in of Fig.~\ref{fig:compare}, showing the clumps of the separate species peaking at different velocities.}
	\label{fig:stylistic}
\end{figure}

In Fig.~\ref{fig:3mm_spectrum}, we found that the HCN 1--0 and 2--1 lines both have outflow features at 500~\kms, whereas the HCO$^+$ 1--0 and 2--1 lines have features at $-450$~\kms. In Figs.~\ref{fig:compare}--\ref{fig:stylistic}, this difference between HCN and HCO$^+$ can be seen more clearly.

To search for further such irregularities, we made PV diagrams of the HCN, HCO$^+$, and HNC emission (Figs.~\ref{fig:pvdiags1}--\ref{fig:pvdiags2} in the online Appendix~\ref{app:pv}). The HCN 1--0 and 2--1 PV diagrams agree well to a first order with the HCN 1--0 and 3--2 PV diagrams presented by \citet{aalto12,aalto15b}. In particular, several spectral features at high velocities and small position offsets are detected. Also in the HCO$^+$ and HNC PV diagrams we detect several such features. The strongest such outflow features are found at approximately 500~\kms, 350~\kms, and $-400$~\kms\ for HCN and 400~\kms\ and $-450$~\kms\ for HCO$^+$. Most of these features are in good agreement with the features in the spectra in Sect.~\ref{sec:results}, but the faintest features cannot be seen in the spectra.

To check for possible contamination of the outflow spectral features, we used the spectral line database Splatalogue\footnote{\url{http://www.splatalogue.net/}} to search for spectral lines with frequencies matching with these clumps. \citet{aalto15b} detected the HCN~3--2 $v_2 = 1$ vibrational line in PdBI observations of Mrk~231. This line blends with the redshifted HCO$^+$ 3--2 outflow line wing, showing up as a feature seen at 400~\kms\ in the HCO$^+$ 3--2 line (see Fig.~1 in \citealt{aalto15b}). The HCN~2--1 $v_2 = 1$ line is, however, not detected in our data (see Fig.~\ref{fig:3mm_spectrum}). This is in agreement with expectations from the HCN~3--2 $v_2 = 1$ line strength: assuming optically thin emission, the HCN~2--1 $v_2 = 1$ should have a line strength around 1.6~mJy\,beam$^{-1}$, about half of the measured flux density at that frequency. Thus, a fraction of the emission detected in the HCO$^+$ 2--1 red wing at around 400~\kms\ should originate in the HCN~2--1 $v_2 = 1$ line emission (this is not coincident with any of the outflow features).
With the exception for the HCN vibrational lines, no likely blend candidates were found in Splatalogue for any of the three bands. The HOC$^+$ 1--0, 2--1, and 3--2 lines should appear at $-1000$~\kms\ in the corresponding HCO$^+$ lines (although the HOC$^+$ 1--0 line is not detected). This velocity is not consistent with any of the suggested clumps, but these lines must be considered when measuring the integrated flux of the blue line wings of HCO$^+$.
Due to the lack of line-blend candidates in the database, we can assume that all features detected in the outflow wings of the HCO$^+$ 1--0 and 2--1 lines at $\varv\gtrsim-820$~\kms, in the HNC line, and in all three HCN lines represent outflowing molecular gas.
PdBI observations of the merging galaxy pair IRAS~F08572+3915 showed similar outflow features in a PV diagram of CO~1--0 emission \citep{cicone14}.

\begin{figure*}[!htb]
    \centering
        $\begin{array}{c@{\hspace{0.0cm}}c}
         \includegraphics{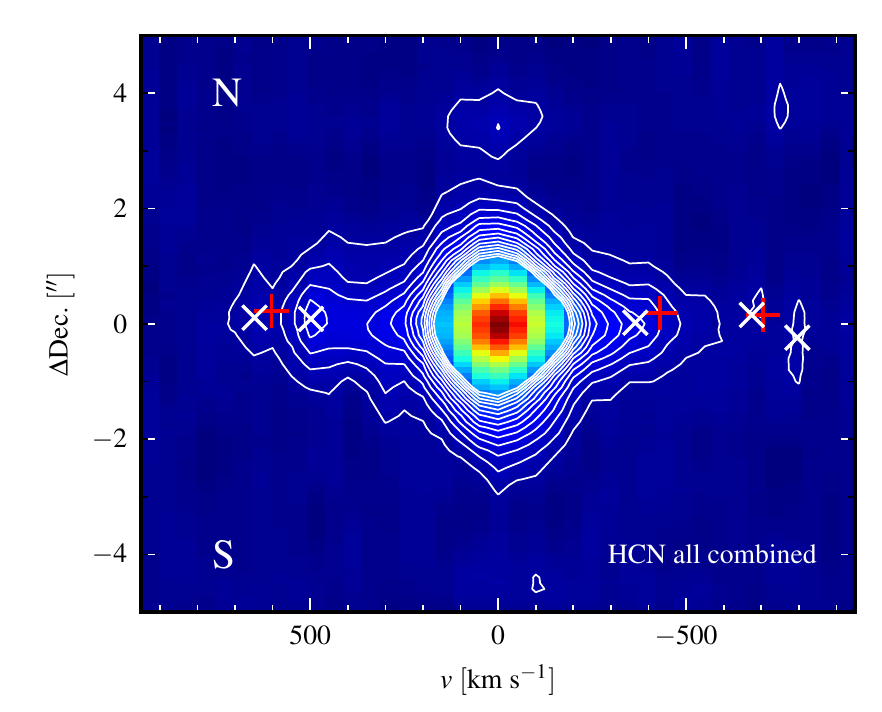} &
         \includegraphics{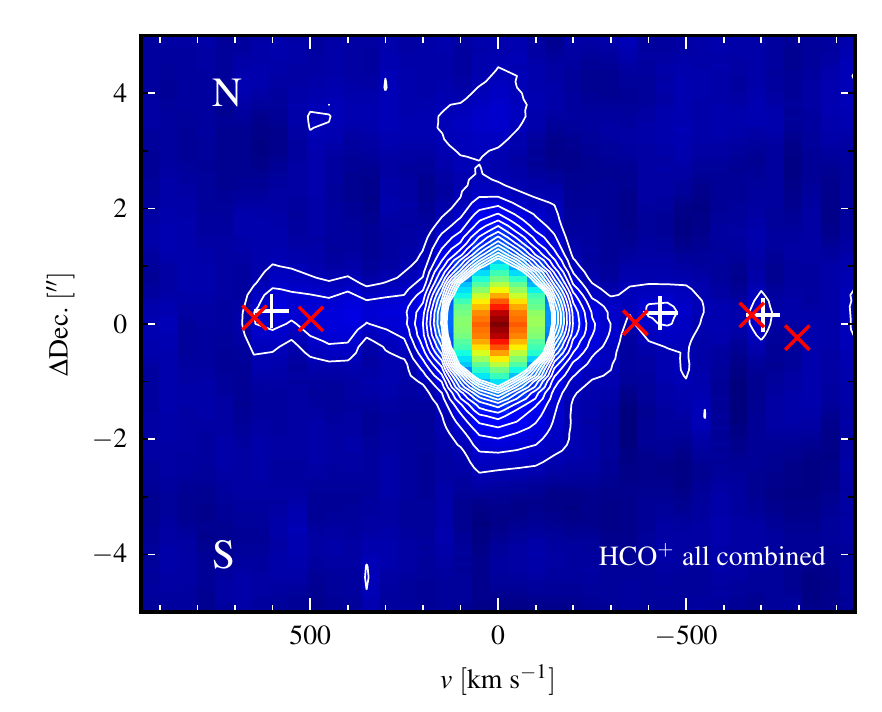} \\
         \includegraphics{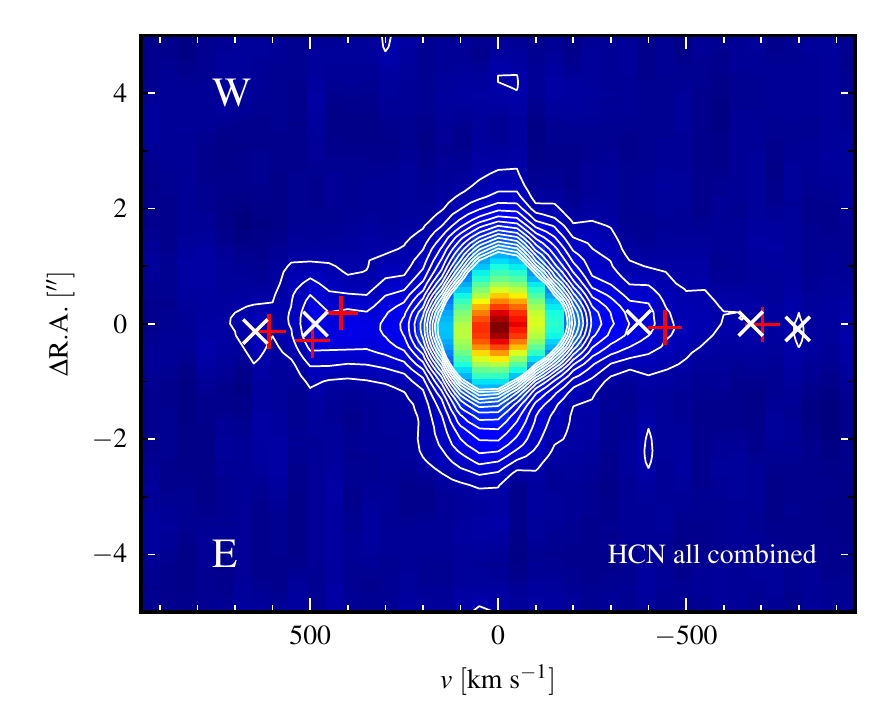} &
         \includegraphics{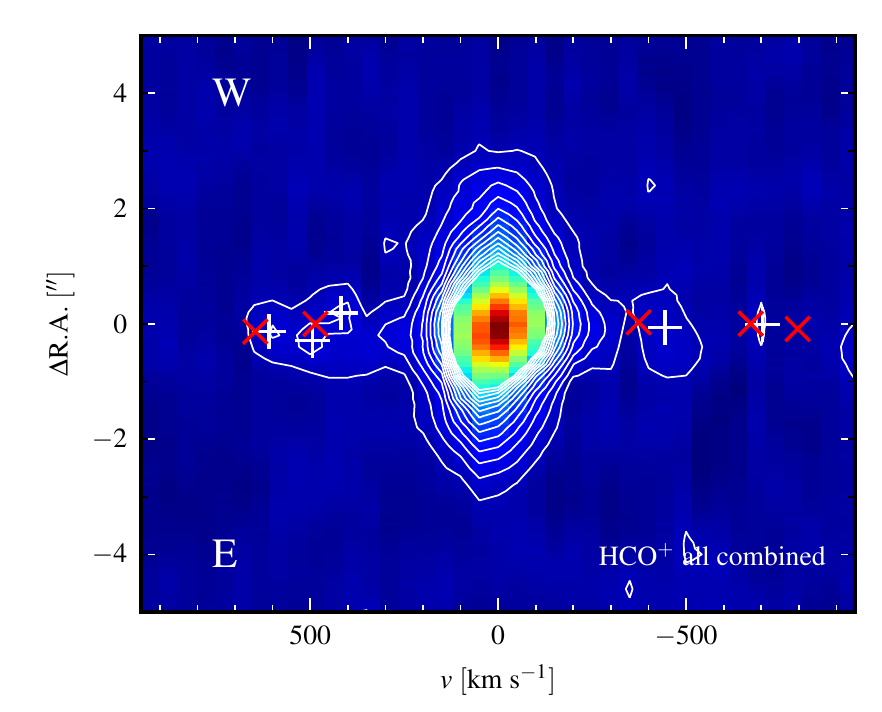} \\
        \end{array}$
    \caption{Stacked PV diagrams of HCN 1--0, 2--1, and 3--2; and HCO$^+$ 1--0 and 2--1 in the combined data (contours start at $2\sigma$, and continue at $2\sigma$ intervals). The HCO$^+$ 3--2 line is not included due to the HCN~3--2 $v_2 = 1$ blend at 400~\kms. In the PV diagrams, the positions of the HCO$^+$ outflow features identified in the stacked PV diagrams are indicated by plus symbols and the HCN features are indicated by crosses. In the HCN PV diagrams, the HCN markers are white and the HCO$^+$ markers are red. In the HCO$^+$ PV diagrams, the HCO$^+$ markers are white and the HCN markers are red.}
    \label{fig:pv_stack}
\end{figure*}

To increase the S/N ratio of these clumps, we also stacked the HCN PV diagrams and the HCO$^+$ PV diagrams. As above, the HCO$^+$~3--2 data were not included due to the HCN~3--2 $v_2 = 1$ blend. In the stacking, the combined A+B-array PV diagrams of the 1--0 and 3--2 lines and the D-array 2--1 PV diagram were used (see Appendix~\ref{app:pv}). The stacking was weighted by the inverse of the rms of each set of observations. In the resulting stacked PV diagrams (Fig.~\ref{fig:pv_stack}), several significant features at high velocity and low position offset are seen. Since we find that the outflow features are amplified by this method it is even less plausible that blends from faint unidentified spectral lines are the cause of this emission.

Like in the spectra (Sect.~\ref{sec:results}), we find that the HCN features appear at different velocities than the HCO$^+$ emission in the PV diagrams. The shifts in position between these features are all within errors of 0\arcsec offset. In Fig.~\ref{fig:pv_stack}, the features are illustrated by plus symbols showing the HCO$^+$ features and crosses indicating the HCN features. For easier comparison, the symbols for both species are shown in all PV diagrams. In the HCN PV diagrams, the HCN markers are white and the HCO$^+$ markers are red. In the HCO$^+$ PV diagrams, the HCO$^+$ markers are white and the HCN markers are red.

\citet{cicone12} found a red outflow component of the CO~1--0 and 2--1 lines in PdBI observations of Mrk~231 ($\varv=527\pm30$~\kms, FWHM = $276\pm71$~\kms). This is consistent with the strongest red clump found in our stacked HCN PV diagram, suggesting that the CO and HCN appear in the same clumps, displaced from the HCO$^+$ features.

The position of the outflow emission is consistent with the line core and continuum emission, also for the high-resolution 3--2 data \citep{aalto15b}. This shows that Mrk~231 has a face-on outflow, in agreement with the face-on disc reported by \citet{bryant96} and \citet{downes98}, and also consistent with the blazar properties as shown by continuum observations (see Sect.~\ref{sec:continuum}).

As seen in Fig.~\ref{fig:compare}, the HNC~1--0 outflow emission \citep{aalto12} also has features at different velocities than HCN and HCO$^+$ (see Fig.~\ref{fig:pvdiags2}), but the S/N level of the HNC 1--0 PV diagram is too low to quantify this.

As a summary of our findings, the HCN outflow emission shows features at around 500~\kms, 350~\kms, and $-400$~\kms, while HCO$^+$ shows fainter features at 600~\kms, 400~\kms, and $-450$~\kms, and HNC has tentative features at 480~\kms, 350~\kms\ and $-340$~\kms. Thus, the outflow emission shows an offset in the peak velocities of these three species. 
The three species HCN, HNC and HCO$^+$ all show features at different velocities in the red outflow. For the blue outflow, the S/N level is lower, but also here tendencies towards differentiation can be seen.

\subsubsection{What causes the chemical differentiation?}

In Mrk~231, the most significant clump of each species appears in the same order in the red and the blue outflow (HCO$^+$ at 600~\kms\ and $-450$~\kms; HCN at 500~\kms\ and $-400$~\kms, and HNC at 350~\kms\ and $-340$~\kms). This apparent symmetry could be an actual effect of the structure and propagation of a shock, or it could be a geometric effect (e.g. an expanding nearly-spherical shell). As the outflow appears to slow down with increasing radius, it suggests that the HCO$^+$ gas is enhanced close to the nucleus and HNC farthest out in the outflow. It should, however, be noted that this clump order is tentative, since a number of fainter clumps also are identified.

Chemical differentiation in molecular outflows has previously been observed towards protostellar sources in the Milky Way \citep{tafalla10,tafalla13}. Like in Mrk~231, the HCO$^+$ component appears at a higher velocity than the HCN component in those protostellar outflows. \citeauthor{tafalla10} note that the outflow component with the highest absolute velocity has a much lower C/O ratio than the normal outflowing gas, suggesting that the high- and low-velocity components have different physical origins.

Since the HCN/HCO$^+$ ratio is enhanced in shocks \citep{mitchell83}, the HCN emission could trace the recently ejected gas which interacts with the surrounding gas in a shock. In the case of a decelerating outflow, and assuming the order of the strongest clumps above, HNC is primarily found in the slower post-shock gas, while HCO$^+$ is found in faster gas closer to the nucleus and before the shock (\citealt{podio14} suggested that HCO$^+$ could be a tracer of pre-shocked gas in protostellar outflows). One complication is, however, the comparison of HCN and HNC with HCO$^+$, where the relative abundance will also be influenced by the O/N abundances in the gas.

High-sensitivity PV diagrams of well-known shock-enhanced species such as SiO, CH$_3$OH, and HNCO will be important to test the hypothesis of shocks and their location in the Mrk~231 molecular outflow. In this scenario we would expect to see enhancements of shock species at velocities where HCN is peaking. To explain why the HNC gas appears at the lowest velocities, post-shock reformation of HNC must be further investigated.

Another possibility that could explain the observed chemical separation is that parts of the molecular outflow could be subject to different amounts of emission from the nuclear region as an effect of the physical structure of the nucleus and jet. In this scenario, it is interesting to note that (again under the assumption of a decelerating outflow) the HCO$^+$ enhancement is closest to the AGN.

\subsubsection{Radiative transfer modelling}

We have performed non-LTE radiative transfer modelling of the outflow emission (red and blue components of HCN, blue components of HCO$^+$) by the use of the RADEX radiative transfer code \citep{vandertak07} to estimate the basic physical properties of the outflowing gas. We test the hypothesis of an outflow consisting of an ensemble of either self-gravitating clouds ($\Delta \varv=\sqrt{GM/R}$) or unbound clouds ($\Delta \varv>>\sqrt{GM/R}$) as suggested by \citet{aalto15b}, following the formalism and assumptions therein. We consider clouds with masses of $10M_\odot$ and $T=50$~K \citep{goldsmith87}; see \citet{aalto15b} for a discussion on these parameters. The self-gravitating cloud model should provide an upper limit of the outflow masses, as the unbound cloud model will result in lower masses. We assume $10\%$ calibration errors on the integrated line fluxes in addition to the statistical errors. The H$_2$ density and the absolute abundances of HCN and HCO$^+$ are set as free parameters. RADEX is used to calculate the line emission from a single clump. Both the modelled line emission from one clump and the observed line emission are then normalised to their respective 1--0 line intensities (to account for the number of clumps), and the difference between these intensities is then minimised to find the best solution. In practice, this means that the number of clumps is the third free parameter.

\begin{figure*}[!htb]
	\centering
	$\begin{array}{c@{\hspace{0.0cm}}c}
	\includegraphics{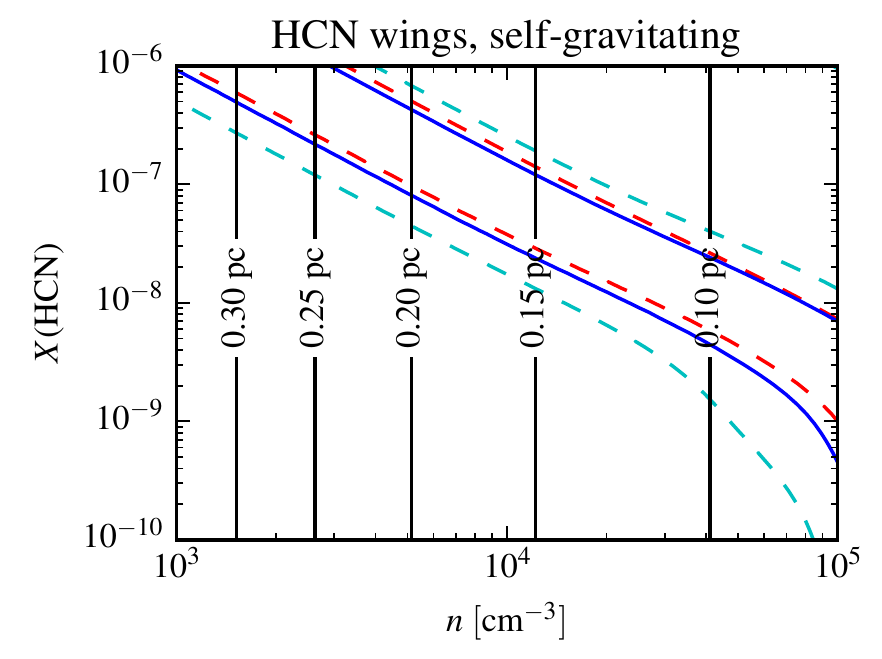} &
	\includegraphics{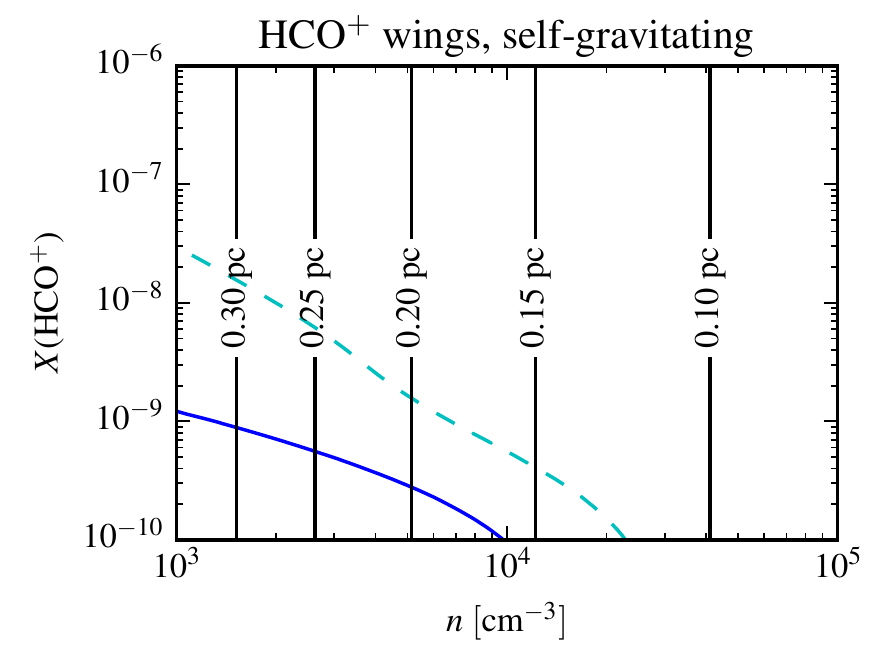} \\
	\includegraphics{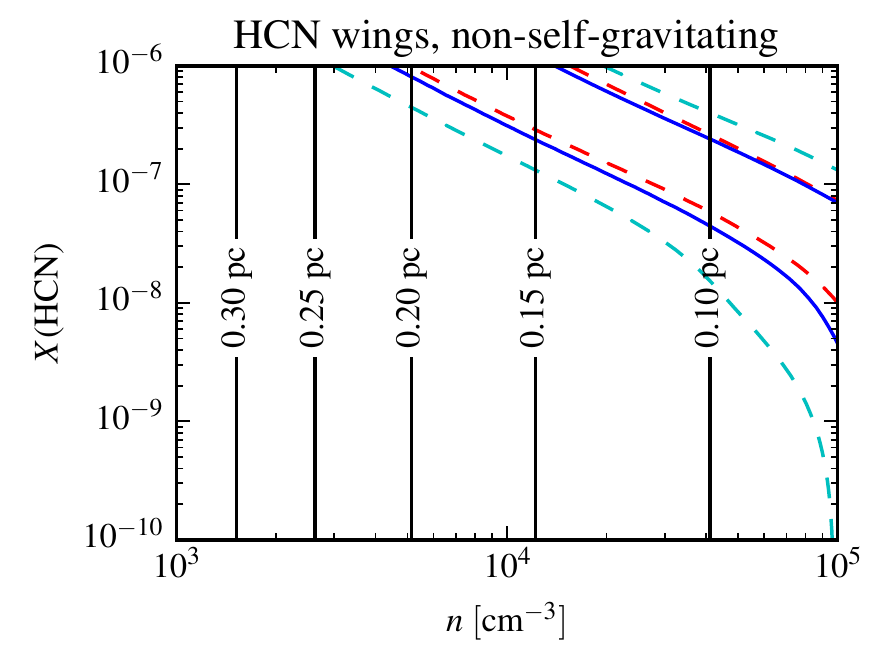} &
	\includegraphics{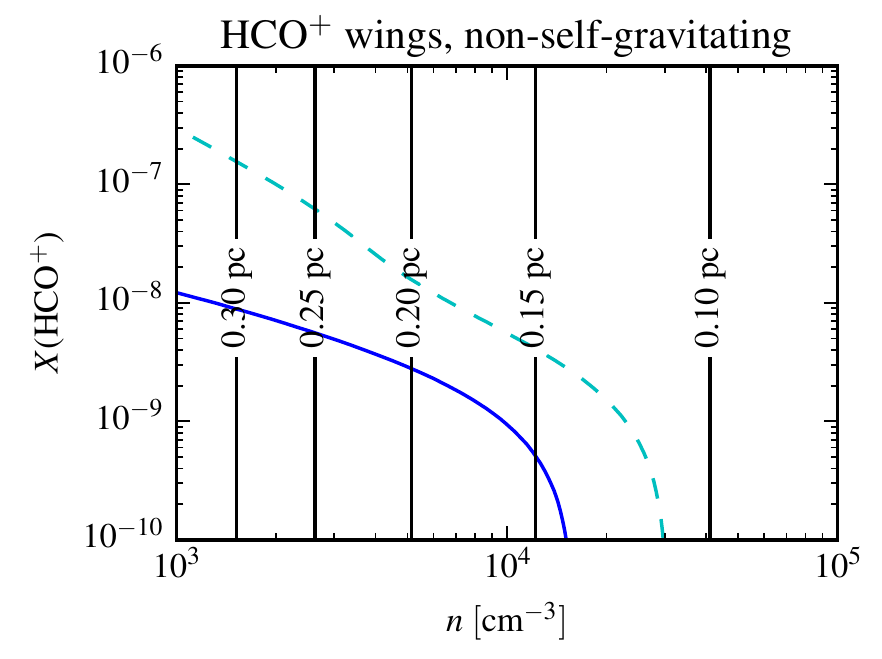} \\
	\end{array}$
	\caption{Least-$\chi^2$ fits to the HCN and HCO$^+$ outflow emission using RADEX radiative-transfer modelling, assuming a clumpy outflow consisting of $10M_\odot$ clouds at $T=50$~K. Red contours correspond to the red wing, and blue/cyan to the blue wing. Contours indicate $2\sigma$ (solid blue) and $3\sigma$ (dashed red and cyan) confidence levels. The black vertical lines show the radii of the individual clouds. For the HCO$^+$, no red wing fits are made due to contamination by vibrational HCN lines.}
	\label{fig:radex}
\end{figure*}

For HCN, we measure the brightness temperature ratios in the red and blue line wings of the line emission towards the central beam, and find that the proportions are similar in the red and blue line wings. The measured line ratios converted to brightness temperature scale are $1:0.75:0.25$ for the transitions 1--0 : 2--1 : 3--2 (cf. columns 6 and 7 in Table~\ref{tab:linefits}). The 1--0/3--2 ratio is slightly lower than the value 0.35 found by \citet{aalto15b}, which is a result of the different methods of extracting this ratio. Our modelling shows that the observed intensities are consistent with cloud sizes of 0.3~pc ($n(\mathrm{H}_2) \approx 2\times10^3$~cm$^{-3}$) assuming $X(\mathrm{HCN})=10^{-6}$, and cloud sizes of 0.1~pc ($n(\mathrm{H}_2) \approx 4\times10^4$~cm$^{-3}$) if lowering the HCN abundance to $X(\mathrm{HCN})=10^{-8}$ (see the upper panels of Fig.~\ref{fig:radex}), which is consistent with the results of \citet{aalto15b}. We reach outflow masses $M_\mathrm{dense} = 3.3\times10^8 M_\odot$ (both wings) for $X(\mathrm{HCN}) = 10^{-6}$, and $M_\mathrm{dense} = 1.4\times10^9 M_\odot$ (both wings) for $X(\mathrm{HCN}) = 10^{-8}$, which are in good agreement with those calculated by \citet{aalto15b}, but they should be seen as an upper limit of the outflow mass since we here assume self-gravitating clumps.

For HCO$^+$, we can only model the blue line wing, since the 3--2 red line wing is blended with the HCN~3--2 $v_2 = 1$ vibrational line. The blue line wings are also blended with HOC$^+$ lines at $-1000$~\kms, so to remove any contribution from this, we only integrate  between $-820$ and $-350$~\kms. The brightness temperature ratios for the blue wing of HCO$^+$ are $1:0.5:0$ for 1--0 : 2--1 : 3--2 (the 3--2 line wing is not significantly detected; cf. column 7 in Table~\ref{tab:linefits}). If we assume the same cloud sizes as derived from the HCN observations, we find that the data are consistent with $X(\mathrm{HCO}^+)\lesssim0.001X(\mathrm{HCN})$ (see Fig.~\ref{fig:radex}). However, considering the clumpy structure and chemical differentiation of the outflow discussed above, this ratio should vary significantly across the outflow. Assuming that the blue and red HCO$^+$ outflows are similar in mass, we reach outflow masses $M_\mathrm{dense} = 2.0\times10^{10} M_\odot$ for $X(\mathrm{HCO}^+) = 10^{-9}$, and $M_\mathrm{dense} = 2.0\times10^{11} M_\odot$ for $X(\mathrm{HCO}^+) = 10^{-10}$. These masses are considerably higher than and inconsistent with those calculated for the total HCN outflow. They are also unrealistically high.

If we instead assume non-self-gravitating clouds for HCN and HCO$^+$ (which corresponds to $\Delta \varv>>\sqrt{GM/R}$; in the lower panels of Fig.~\ref{fig:radex} we have assumed $\Delta \varv=10\sqrt{GM/R}$), we reach lower outflow masses, in agreement with \citet{aalto15b}. Compared to the self-gravitating clouds, the outflow masses calculated for both the HCN and HCO$^+$ components are approximately a factor 5 lower when assuming $\Delta \varv=10\sqrt{GM/R}$ but keeping the HCN and HCO$^+$ abundances unchanged. Thus, when assuming identical physical properties of the HCN and HCO$^+$ outflows and non-self-gravitating clouds, the HCN and HCO$^+$ models produce strongly contradictory outflow masses, just as when assuming self-gravitating clouds. The modelled HCO$^+$ outflow masses are also so high that the model assumptions must be incorrect (e.g. too high temperature), while those same assumptions appear to work for HCN. Consequently, the models support the previous conclusion of a chemical differentiation between HCN and HCO$^+$ in the molecular outflow.

\subsection{Variable continuum}
\label{sec:continuum}

The high variability detected in the 3~mm, 1~mm, and 1.5~cm (20~GHz) continuum emission of Mrk~231 on timescales of a few years indicates that Mrk~231 is a blazar.

The time coverage of our 3~mm observations is unfortunately not good
enough for a reliable modelling of the flaring event of Mrk~231. We
can, however, discuss the consistency between our results and those at
other frequencies \citep{aalto12, reynolds13, aalto15b} in the frame
of the standard jet model \citep[e.g.][]{marscher80, blandford79}.

The light curve at 20~GHz reported in \cite{reynolds13} peaks on
early March 2013, when the flux density roughly doubled, compared to
its value on early January 2013. Our flux density at 3~mm in January
2013 is, however, already twice that reported by \cite{aalto12}. A
similar result is found at 1~mm, as already mentioned in Sect.~\ref{sec:results}: the flux density, compared to that in 2012, doubled
on February 2013 (i.e., well before the peak intensity was reached at
20~GHz, on March 2013).
The flux density at mm wavelengths thus seems to have varied by a
factor $\sim$2 well before the peak intensity at 20~GHz was reached.
This is indicative of opacity effects in the jet
\citep[e.g.][]{blandford79}, which at lower frequencies would make the
jet opaque at shorter distances to its base. Hence, flaring activity
propagating downstream of the jet would be detectable earlier at
higher frequencies, and later at lower frequencies.

\section{Conclusions}
\label{sec:conclusions}

In this paper we present high-resolution observations of the outflow of the ULIRG Mrk~231. The galaxy has a very complex outflow structure, which is barely resolved in the PdBI observations. We have examined the chemical composition of this gas, and identify the following signs of chemical differentiation in the outflow:

\begin{enumerate}
\item The HCN line core emission of Mrk~231 is broader than the HCO$^+$ and HNC line core emission. This could represent HCN emission at forbidden velocities, but is more likely caused by optical-depth broadening of the HCN line.
\item The HCN, HCO$^+$, and HNC emission line wings all show features at different velocities, suggesting a clumpy, chemically differentiated outflow.
\item By radiative transfer modelling we show that while the HCN emission is consistent with a clumpy outflow, the same model properties applied to the HCO$^+$ emission generates unrealistically high outflow masses, which are also inconsistent with the HCN modelling results. This suggests that the HCN and HCO$^+$ emission originates in structures with different physical properties.
\end{enumerate}

The chemical differentiation of HCN and HCO$^+$ can be explained by an enhancement of the HCN/HCO$^+$ abundance ratio in shocked parts of the outflows \citep{mitchell83}. The HCO$^+$ emission, at the highest velocities, traces pre-shock gas (possibly at lower temperature than the HCN). The HCN emission traces recently ejected gas interacting in a shock with the surrounding gas. The HNC appears at the lowest velocities, and traces the braked gas after the shock.

We also report that the 3~mm continuum flux of Mrk~231 was significantly enhanced by the radio flare event reported by \citet{reynolds13}, indicating that this galaxy is a blazar.

Future observations using NOEMA could be used to achieve data with higher sensitivity and resolution, which is required to further investigate the nature of the complex outflow structure of this interesting object. This would improve our understanding of the origin of the outflows from ULIRGs. To better study the importance of shocked gas, such observations should not only target HCN and HCO$^+$, but also shock tracers such as SiO, CH$_3$OH, and HNCO.

\begin{acknowledgements}
      We thank the IRAM PdBI staff for excellent support. This research was supported by an appointment to the NASA Postdoctoral Program at the NASA Goddard Space Flight Center to J.E.L., administered by Oak Ridge Associated Universities through a contract with NASA. S.A. thanks the Swedish National Science Council for grant support. We also thank the referee Chiara Feruglio for comments and suggestions which significantly contributed to improve the quality of the manuscript.
\end{acknowledgements}

\bibliographystyle{aa}
\bibliography{mrk231_paper}
\clearpage
\onecolumn

\clearpage
\begin{appendix}
\section{PV diagrams}
\label{app:pv}

We here show the PV diagrams for each individual HCN and HCO$^+$ line reported for the first time in this work. They were all obtained by a cut through the continuum peak, either in north-south or east-west direction. All contours start at $2\sigma$ and continue with $2\sigma$ spacing (see Table~\ref{tab:obs} for rms levels).

\begin{figure*}[!htb]
	\centering
	$\begin{array}{c@{\hspace{0.0cm}}c}
	\includegraphics{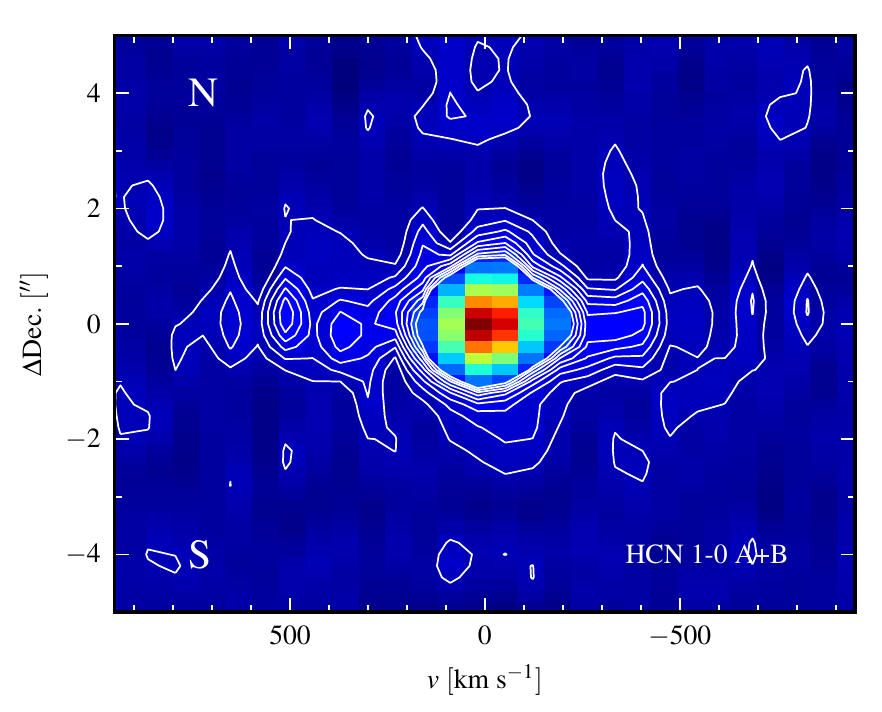} &
	\includegraphics{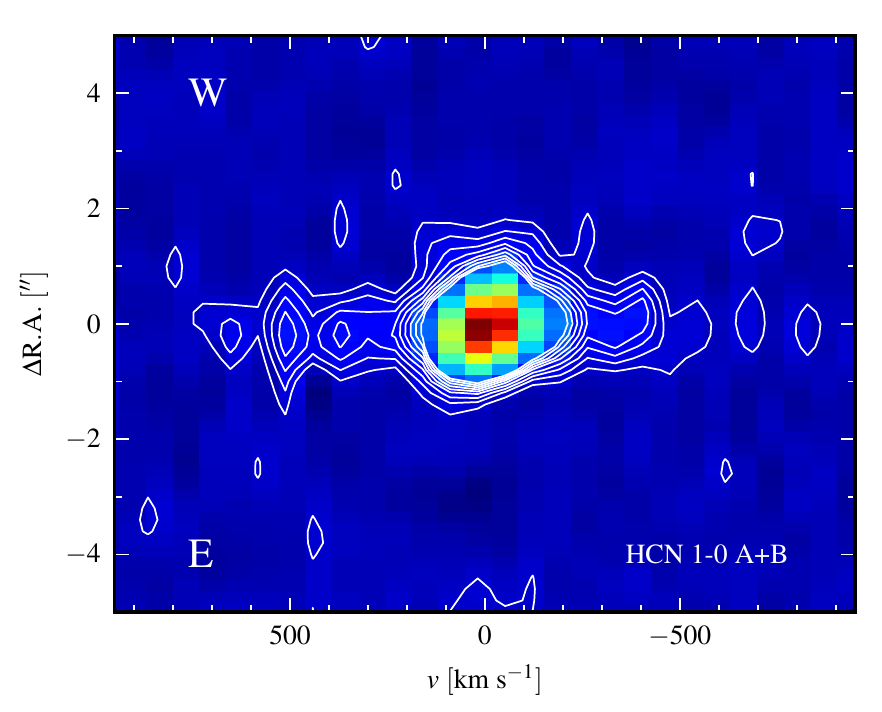} \\
	\includegraphics{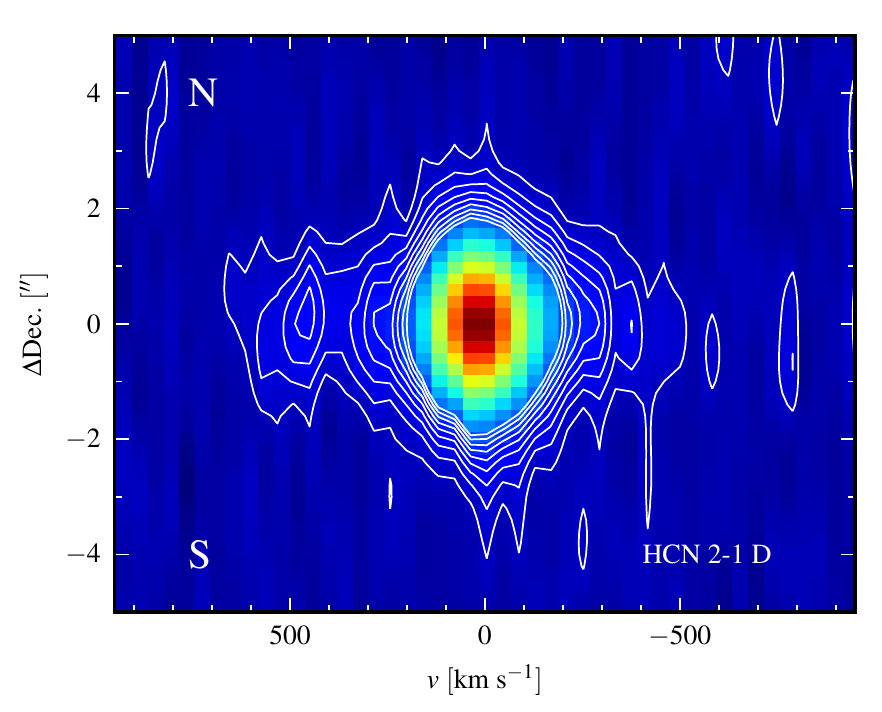} &
	\includegraphics{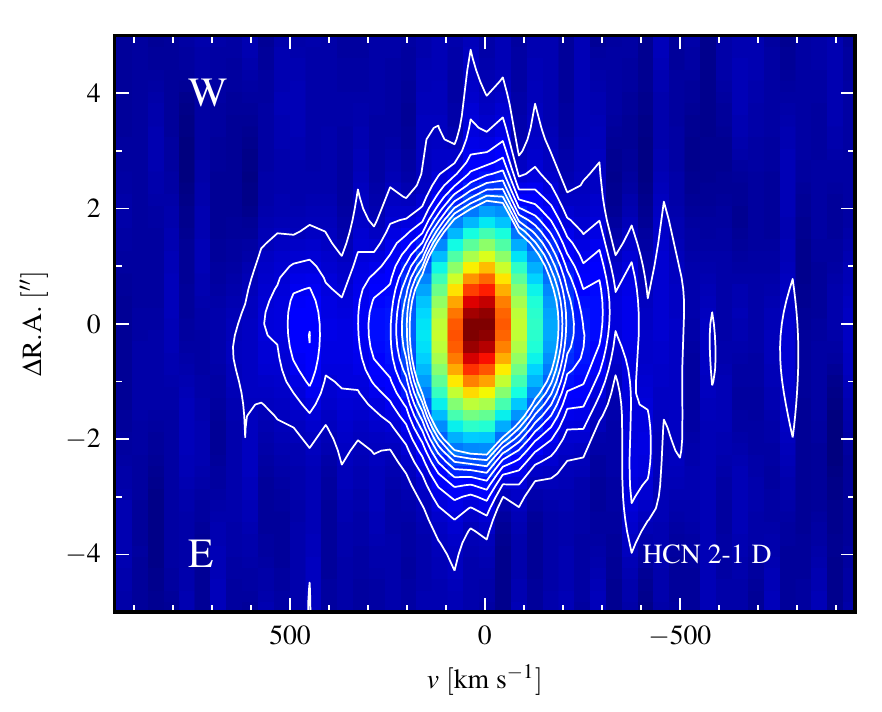} \\
	
	\end{array}$
	\caption{PV diagrams of individual HCN transitions (contours at $2\sigma$). The HCN 1--0 PV diagram uses a combination of PdBI A-array and B-array observations. The HCN 2--1 PV diagram is made from PdBI D-array observations.}
	\label{fig:pvdiags1}
\end{figure*}

\begin{figure*}[!htb]
	\centering
	$\begin{array}{c@{\hspace{0.0cm}}c}
	\includegraphics{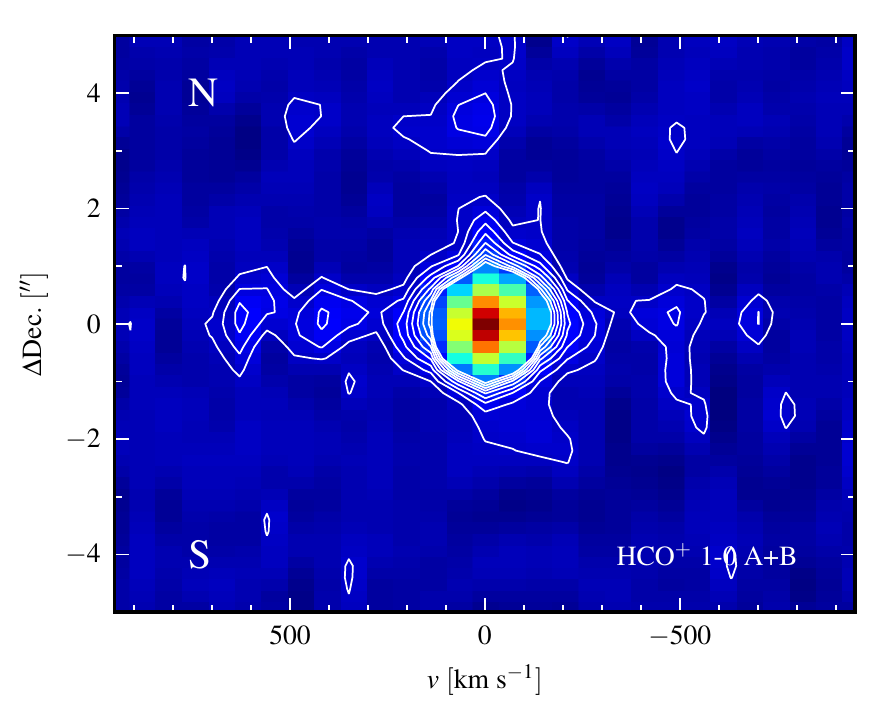} &
	\includegraphics{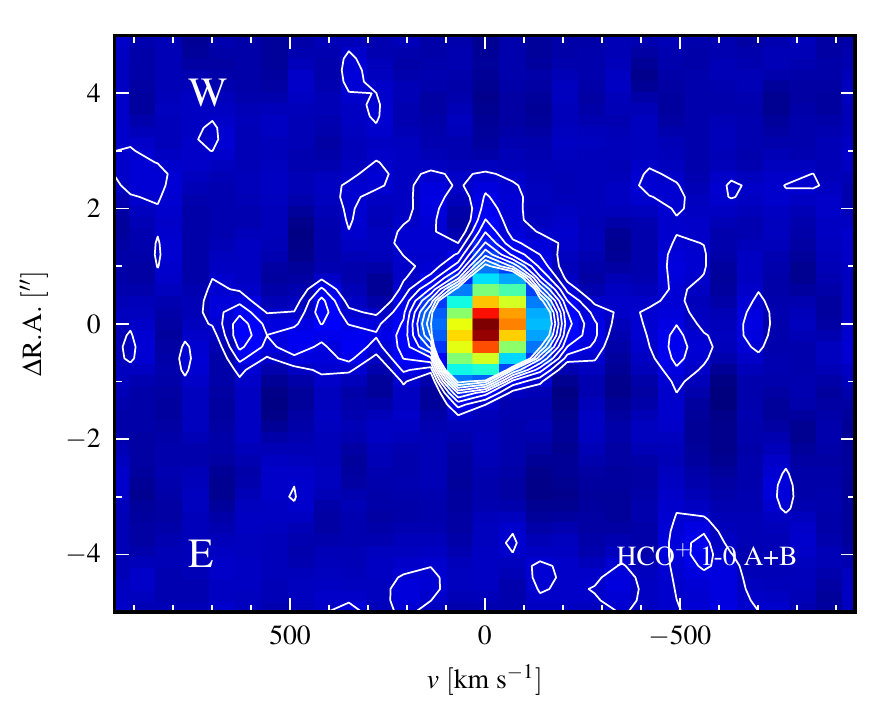} \\
	\includegraphics{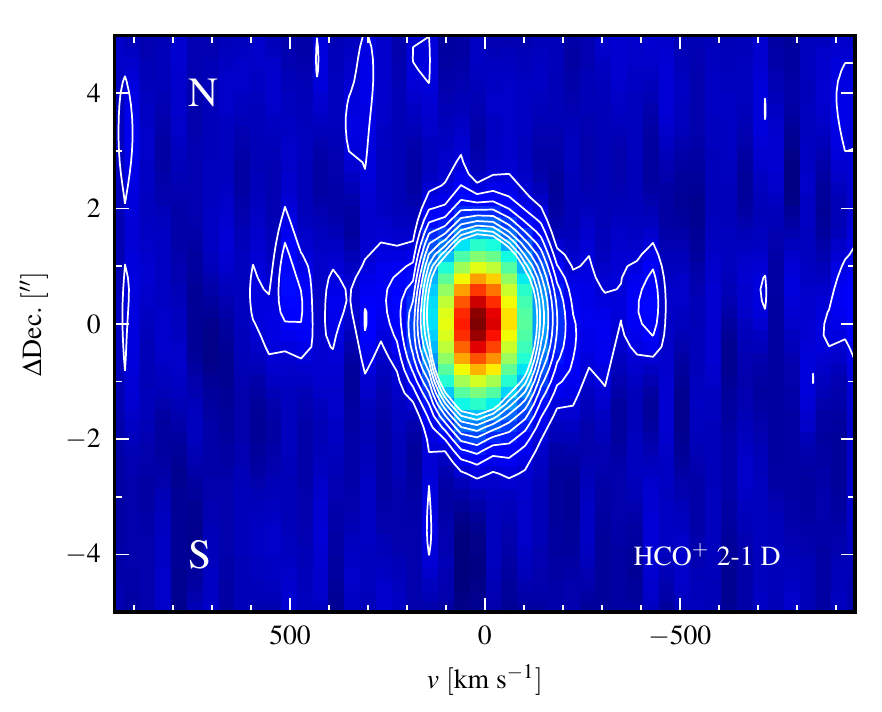} &
	\includegraphics{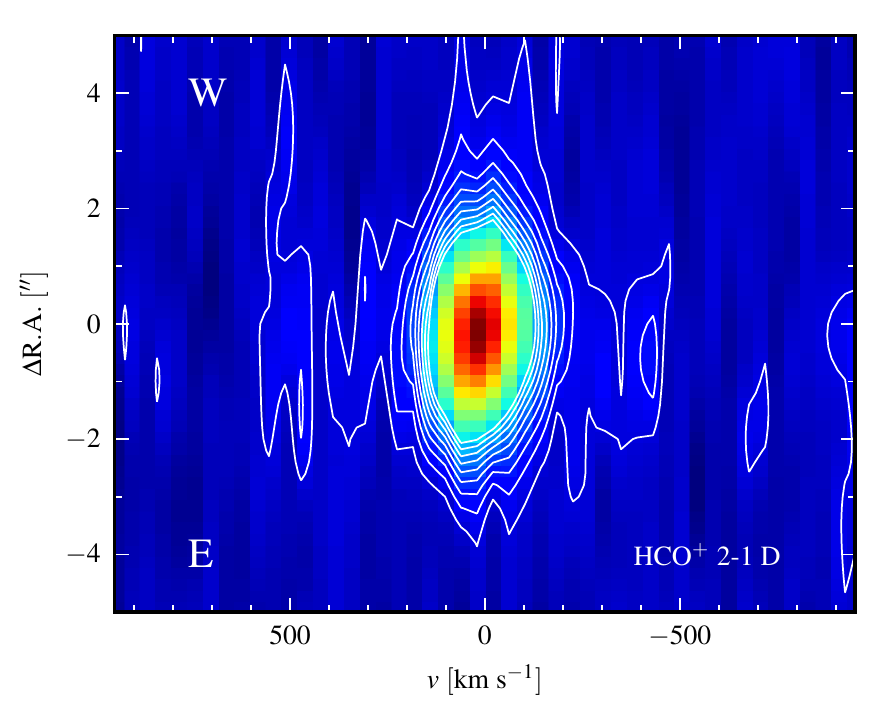} \\
	\includegraphics{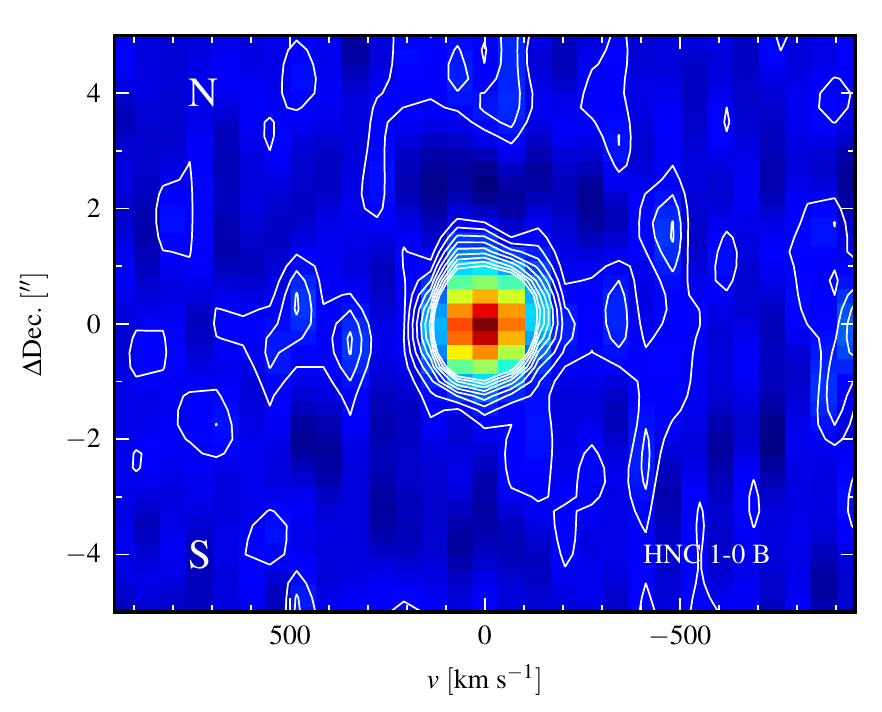} &
	\includegraphics{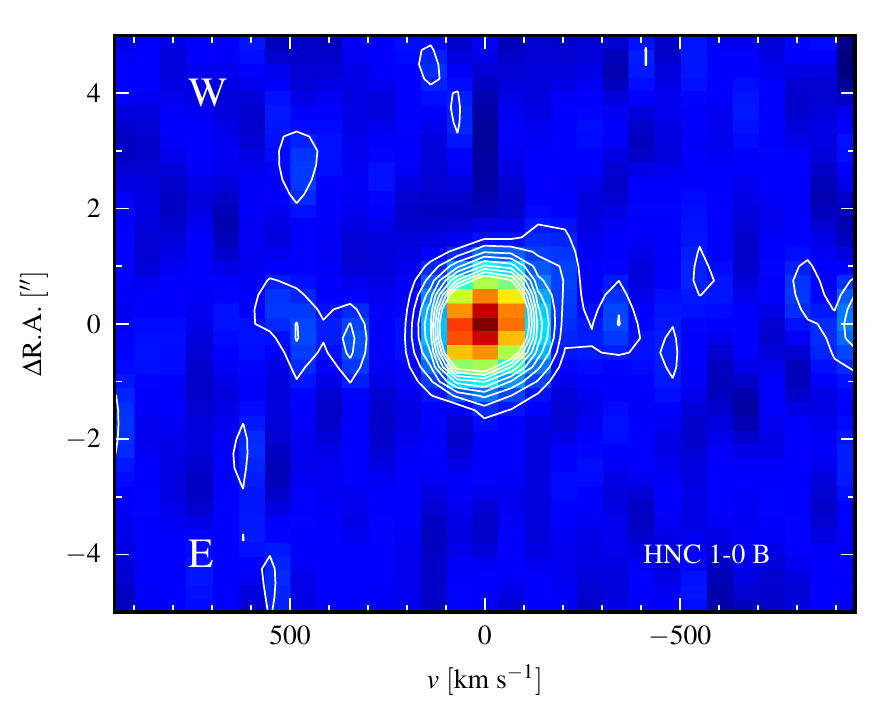} \\

	\end{array}$
	\caption{PV diagrams of individual HCO$^+$ and HNC transitions (contours at $2\sigma$). The HCO$^+$~1--0 PV diagram uses a combination of PdBI A-array and B-array observations. The HCO$^+$~2--1 PV diagram is made from PdBI D-array observations. The HNC~1--0 PV diagram is made from PdBI B-array observations.}
	\label{fig:pvdiags2}
\end{figure*}

\end{appendix}

\end{document}